\documentclass[useAMS,usenatbib]{mn2e}

\usepackage{txfonts}
\usepackage[dvips]{graphicx}
\usepackage{epsfig}
\usepackage{float}
\usepackage{subfigure}

\def\lsim{\mathrel{\hbox{\rlap{\hbox{\lower4pt\hbox{$\sim$}}}\hbox{$<$}}}}
\def\gsim{\mathrel{\hbox{\rlap{\hbox{\lower4pt\hbox{$\sim$}}}\hbox{$>$}}}}

\newcommand{\rvoid} {\mathrm{R_{void}}}
\newcommand{\rmpc}  {\mathrm{R_{void}/h^{-1}\,Mpc}}

\title[Large scale environment around voids]%
{Clues on void evolution I:\\ Large scale galaxy distributions around voids}

\author[Ceccarelli et al.]{
\parbox[t]{\textwidth}
{ L. Ceccarelli$^{1,2}$\thanks{E-mail: ceccarelli@oac.uncor.edu},
  D. Paz$^{1,2}$,
  M. Lares$^{1,2}$,
  N. Padilla$^{3}$
  \& D. Garc\'{\i}a Lambas$^{1,2}$
}
\vspace*{6pt}\\
$^1$ Instituto de Astronom\'\i a Te\'orica y Experimental, 
     UNC-CONICET, C\'ordoba Argentina. \\
$^2$ Observatorio Astron\'omico de C\'ordoba, UNC, Argentina. \\
$^3$ Departamento de Astronom\'\i a y Astrof\'{\i}sica, Pontificia
     Universidad Cat\'olica de Chile, Santiago, Chile.\\
}

\begin{document}

\date{\today}

\maketitle

\begin{abstract}
We perform a statistical study focused on void environments.
We examine galaxy density profiles around voids in the SDSS, finding a
correlation between void--centric distance to the shell of maximum
density and void radius when a maximum in overdensity exists.
We analyze voids with and without a surrounding over-dense shell in
the SDSS.
We find that small voids are more frequently surrounded by over-dense
shells whereas the radial galaxy density profile of large voids tends
to rise smoothly towards the mean galaxy density.
We analyse the fraction of voids surrounded by overdense shells
finding a continuous trend with void radius.
The differences between voids with and without an overdense shell around
them can be understood in terms of whether the voids are, on average, in the
process of collapsing or continuing their expansion, respectively, in 
agreement with previous theoretical expectations.
We use numerical simulations coupled to semi-analytic models of galaxy
formation in order to test and interpret our results.
The very good agreement between the mock catalog results and the
observations provides additional support to the viability of a $\Lambda$CDM
model to reproduce the large scale structure of the universe as
defined by the void network, in a way which has not been analysed
previously. 
%
\end{abstract} 


\section{Introduction} \label{S_intro}


  \begin{figure*}
  \includegraphics[width=\textwidth]{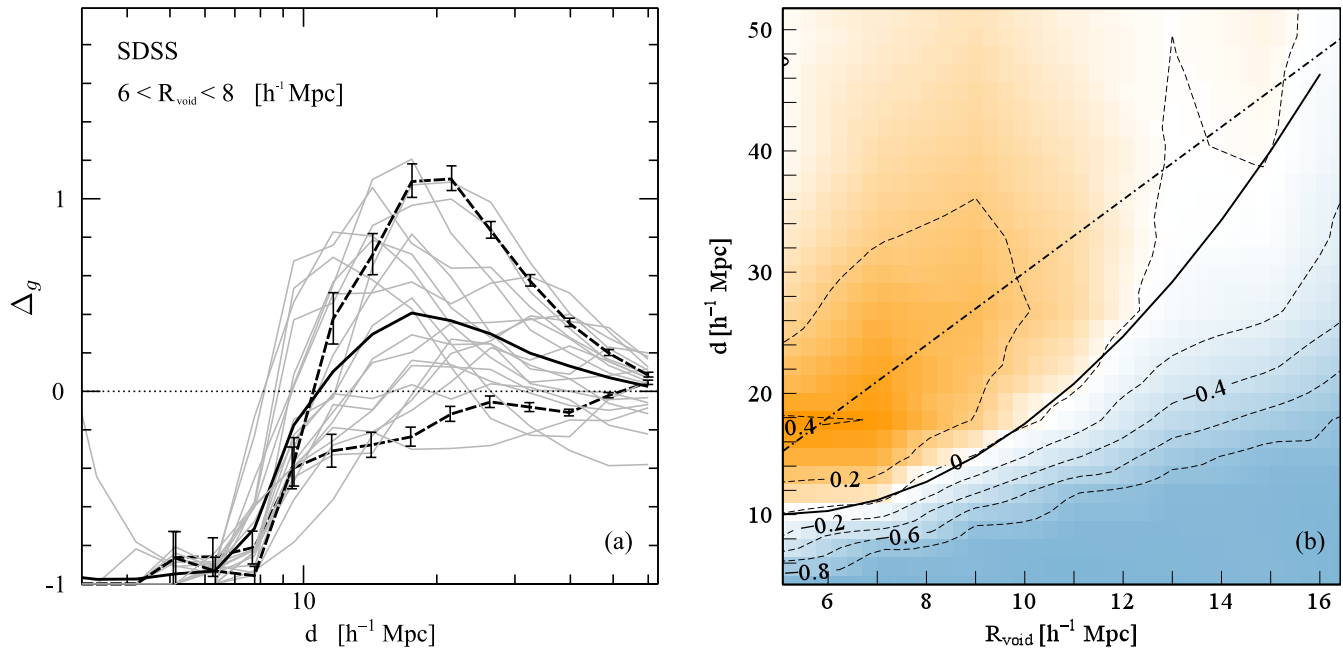}%
  \caption{
	\textbf{
 \textit{Left panel:}}
     Integrated galaxy density ($\Delta_g$) as a function of distance to the
     void centre for individual voids in SDSS with radii in the
     range 6-8 h$^{-1}$Mpc (gray lines). The thick black solid line indicates the mean
     density contrast profile of all voids.
     The thick dashed lines show two different profiles:  
     with and without a noticeable maximum in the density profile
     and the error bars represent the corresponding poisson errors.
	\textbf{
 \textit{Right panel:}}
     Contour lines of mean density contrast as a function of void
     radius and distance to the void centre d in SDSS. 
     Orange colours represent positive densities
     (increasing from light to dark) and cyan correspond to negative
     densities.  Dashed lines represent iso-density contours
     ($\Delta_g$ = -0.8, -0.6, -0.4, -0.2, 0, 0.2, 0.4) as is labelled
     in the figure.  The solid line represents an approximate fit to
     the zero density contour line. 
     The dot-dashed lines indicates the relation 3$\times \rvoid =$ d.
}
  \label{F_SDSS_dens} 
  \end{figure*}



Although the large scale galaxy and matter distributions are dominated by
different structures such as groups, clusters, filaments, walls etc.,
most of the volume of the Universe is occupied by voids: large
underdense regions are the prominent structures at large scales.

These large scale underdensities have been identified and analyzed in
numerical simulations and in galaxy catalogues \citep{hoffman_shaham,
hausman, fillmore_goldreich, icke_1984, bertschinger, pellegrini,
kauffmann_voids_1991, blumenthal_largest_1992, slezak_1993,
el-ad_1997, el-ad_piran_1997, el-ad_piran_2000,
aikio_maehoenen_1998, muller_2000, plionis_basilakos_2002,
hoyle_vogeley_2002, sheth_hierarchy_2004, hoyle_voids_2004,
ceccarelli_voids_2006, tikhonov_2006, patiri_2006,
furlanetto_piran_2006, Neyrinck_2008, aragon-calvo_unfolding_2010, pan2012, sutter_public_2012}.
 Despite the fact that the agreement between void finders is weaker for small 
and medium scale voids
to a first approximation their more prominent properties are similar. 
In general, similar properties of voids are found regardless of the
diversity of identification methods and galaxy sample properties. 
For a comparison of the different techniques adopted see
\citet{colberg_aspen-amsterdam_2008}.

It is also important to stress the fact that 
galaxies and haloes trace the void distribution in a similar fashion.  
\citet{padilla_spatial_2005} have studied voids defined by the spatial 
distribution of haloes and galaxies finding that they have comparable general 
statistical and dynamical properties, such as abundances, correlation functions 
and velocity fields.

Moreover, the statistics of void and matter distributions are strongly
related \citep{white79} and therefore voids can provide simple and
useful information on the clustering pattern, giving clues on the
formation and evolution of overdense structures. 
Observations of the large scale structure traced by voids can also be
used to constrain cosmological models
\citep[e.g.][]{peebles_void_2001, park_challenge_2012,
   kolokotronis_supercluster_2002, colberg_voids_2005,
lavaux_precision_2010, bos_darkness_2012, biswas_voids_2010,
benson_galaxy_2003, park_challenge_2012, bos_less_2012,
hernandez-monteagudo_signature_2012, clampitt_voids_2012,
sutter_first_2012} and to shed light on the mechanisms of 
galaxy evolution and its dependence on the large scale 
environment \citep{lietzen_environments_2012, hahn_properties_2007, 
hahn_evolution_2007, ceccarelli_low_2012, ceccarelli_large-scale_2008}.

Voids in the galaxy distribution, in a first approximation, can 
be described as simple underdense regions which have nearly spherical %
shapes and isotropic expansion motions \citep{icke_1984, 
weygaert_bertschinger_1996, padilla_spatial_2005, ceccarelli_voids_2006}.  
Nevertheless, in more detailed analyses it became clear
that as voids are not isolated structures but rather part of an intricate
network, their structure and dynamics are more complicated
\citep{bertschinger, melott_shandarin_1990, mathis_white_2002,
colberg_etal_2005, Shandarin_etal_2006, platen_etal_2008,
aragon-calvo_hierarchical_2012}.  The formation and evolution of voids
is strongly affected by their surrounding large scale environment
\citep{sheth_hierarchy_2004}.  The void distribution also evolves
accordingly, as matter collapses to form structures and galaxies %
dissipate from voids, making a supercluster-void network
\citep{frisch_evolution_1995, einasto_supercluster_1997,
einasto_multimodality_2012}.  In order to deepen our understanding of
the nature of voids and the evolution of their properties, it is
crucial to take into account the large scale structure where they are
embedded \citep{more_a_do_about_nothing}.

Essentially, the hierarchy of voids arises by the assembly of matter
in the growing nearby structures.  \citet{sheth_hierarchy_2004}
suggest that while some voids remain as underdense regions, other
voids fall in on themselves due to the collapse of dense structures
surrounding them.
According to this scenario, void evolution exhibits two opposite
processes, expansion and collapse, being the dominant process
determined by the global density around the voids. The distinction
between these two types of void behavior depends on the surrounding
environment.
It is expected that the large underdense regions with surrounding
overdense shells will undergo a ''void-in-cloud'' evolution mode.
These voids are likely to be squeezed as the surrounding structures
tend to collapse onto them.

On the other hand, voids in an environment more similar to the global
background density will expand and remain as underdense regions
following a so called ''void-in-void'' mode.
In this scenario, it could be expected that largest voids at present
remain stable, and thus are unlikely to be surrounded by overdense
regions that are massive enough to produce a contraction.
However, many of the smallest voids at present may show surrounding
overdense shells.
Inspired in this schematic scenario we analyze the void-size
dependence of the relative population of voids embedded in low density
and overdense regions.

This paper is organized as follows:
in Section \ref{S_data} we describe the galaxy and mock
catalogs and the corresponding void catalogues.
In Section \ref{S_modes} we provide the analysis of the environments
of voids in the SDSS.
A comparison of observational results to the numerical simulation and
the mock catalogue is given in Section \ref{S_teo}.
Finally, we discuss our results in Section \ref{S_concl}.


\section{Data } \label{S_data}

\subsection{Galaxy catalogue and void sample}

The observational data used in this work were extracted from the Main
Galaxy Sample \citep{strauss_spectroscopic_2002} of the Sloan Digital
Sky Survey data release 7
\citep[\textsc{SDSS-DR7,}][]{abazajian_seventh_2009}.
The SDSS contains CCD imaging data in five photometric bands
\citep[UGRIZ,][]{fukugita_sloan_1996, smith_ugriz_2002}.
The \textsc{SDSS-DR7} spectroscopic catalogue comprises in this
release 929,555 galaxies with a limiting magnitude of
\mbox{$r~\leq~17.77$ mag}.

The voids in the galaxy distributions are identified using volume complete
samples.
We adopt a limiting redshift $z=0.08$ and maximum absolute magnitude
in the $r-band$ $M_r=-19.2$.
The limiting redshift of the sample is chosen on the basis of a
compromise between the quality of the void sample, determined by the
dilution of the sample of galaxies, and the number of voids, required
to be large in order to achieve statistically significant results.

We have explored the effects of shot noise on the 
void identification in both real data and simulations in \citet{ceccarelli_voids_2006}.
We found that for 
low-density galaxy samples, the identification of %
small voids ($R_{void} <$ 10 h$^{-1}$ Mpc) is limited by shot noise, 
however we can identify voids as small as these in denser galaxy samples %
that can be obtained by restricting the analysis to low redshifts ($z \lessapprox 0.1$ in SDSS).%

Our sample is redshift limited to have a large enough galaxy density 
to avoid the effects of shot noise on small voids (down to 5 h$^{-1}$ Mpc), 
but also a large enough
volume to provide a statistically significant number of objects. 
We apply the void finding algorithm described in
\citet{padilla_spatial_2005} and \citet{ceccarelli_voids_2006} to this
volume limited sample. In order to prevent effects of
survey geometry and redshift limits on the process of void identification we avoid including voids %
near the survey edges.
The algorithm identifies the largest spherical regions where the
overall density contrast is at most $\Delta=-0.9$, and are not contained in
any other sphere satisfying the same condition.
According to this procedure, a void is located at the centre of an
underdense sphere and has a scale size equal to the sphere
radius.
We obtain 131 voids in the SDSS sample
with radii ranging from 5 h$^{-1}$Mpc to 22 h$^{-1}$Mpc.

\subsection{Large scale numerical simulation}
\label{ss:sim_mock}

We use the snapshot corresponding to $z=0.0$ of the 
Millennium simulation \citep{springel_simulations_2005,
lemson_halo_2006} and the associated semi-analytic model of galaxy
formation by \citet{bower_flip_2008}.
The Millennium simulation utilizes a $\Lambda$CDM cosmological model
with \mbox{$\Omega_{\Lambda}$ = 0.75}, \mbox{$\Omega_M$ = 0.25},
\mbox{$\Omega_b$ = 0.045}, \mbox{h = 0.73}, \mbox{n = 1} and
\mbox{$\sigma_8$ = 0.9} based on WMAP observations
\citep{spergel_first-year_2003} and 2dF Galaxy Redshift Survey
\citep{colless_2df_2001}.
The simulation follows the evolution of $2160^3$ particles, each with
8.6 $\times\, 10^8\, \mathrm h^{-1}M_{\odot}$ through a comoving box
of side \mbox{500 $\mathrm Mpc$}.
The dark matter haloes of the Millennium simulation were used to
follow the simulated growth of galaxies by implementing a
semi-analytic model of galaxy formation
\citep[\textsc{GALFORM},][]{bower_flip_2008}, which generates a
population of galaxies within the simulation box.
We used the full simulation box to identify voids and study the
effects of dilution and redshift space distortion in the sample of
voids.
The semi-analytic galaxy catalogue from the simulation box comprises
2783 voids.

We use a mock catalogue, constructed by selecting galaxies within the
simulation box.
The mock catalogue is constructed by first positioning an observer in
a random position within the numerical simulation box, and then
reproducing the selection function and angular mask of the SDSS from
this position.
This results in a mock catalogue of galaxies of similar properties and
observational biases to those of the  SDSS catalogue.
Positions in real space and peculiar velocities are available to test
possible projection biases and to quantify the effects of redshift
space distortions.
The mock catalogue also provides information on SDSS photometric
magnitudes, star formation rates and total stellar masses, based on
computations from the semi--analytic model of galaxy formation
\citep{bower_flip_2008}.
This mock catalogue will be used in this work to calibrate our
statistical methods, to interpret the data, and to detect any
systematic biases in our procedures.
In order to do this, we will treat the mock catalogue in exactly the
same way as the real data. 
We found 113 mock voids, with radii ranging from \mbox{5 h$^{-1}$Mpc}
to \mbox{23.75 h$^{-1}$Mpc}.


 \begin{figure}
 \includegraphics[width=84mm]{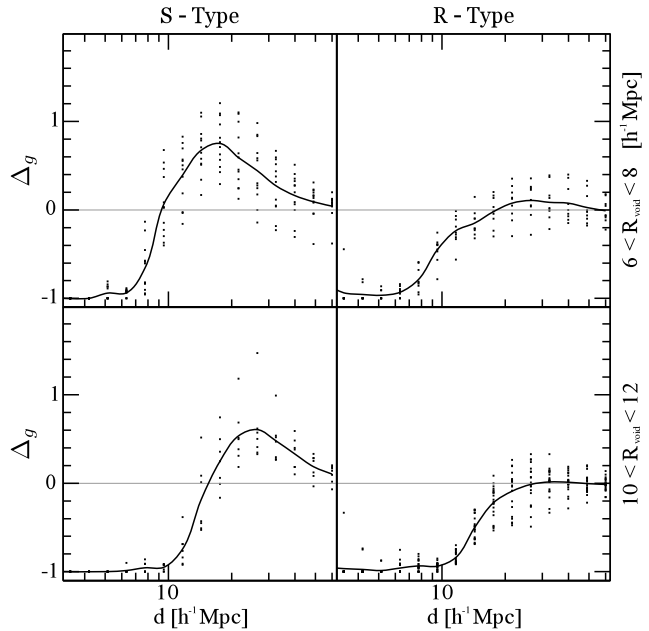}
 \caption{Integrated density contrast profiles around voids of radii
 in the range 6-8 h$^{-1}$Mpc (upper panels) and
 10-12 h$^{-1}$Mpc in SDSS (lower panels).
 The profiles in the left panels correspond to
 voids surrounded by overdense shells, i.e., S-Type voids.
 Right panels are for R-Type voids, showing a non--decreasing
 or ''rising'' profile.
 Solid thick lines correspond to the mean density contrast profiles
 and black dots show the values of density contrast
 for each individual void as a function of distance.
 Dotted lines indicate the mean galaxy
 density, i.e., zero density contrast.}
 \label{F_SDSS_prof} 
 \end{figure}

 \begin{figure}
 \includegraphics[width=84mm]{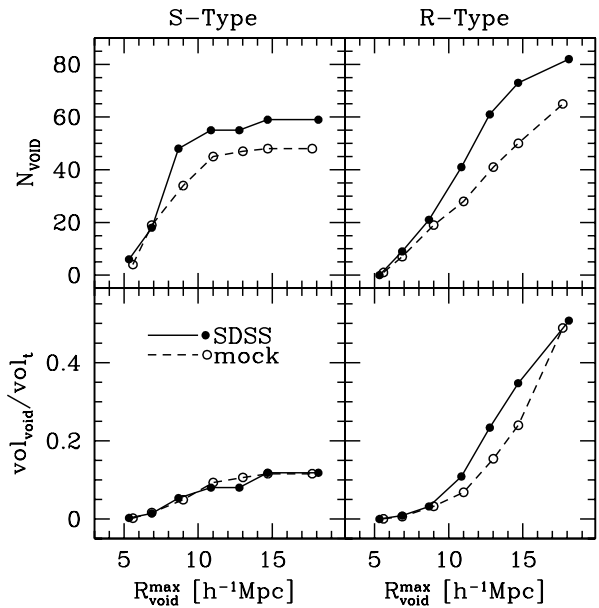}
 \caption{
	\textbf{
 \textit{Upper panels:}}
 Integrated number of voids as a function of radius in SDSS (solid lines, 
 filled circles) and in the mock (dashed lines, open circles) catalogues, 
 for voids with (S-type) and without (R-type) a surrounding overdense shell 
 (left and right panels respectively).
	\textbf{
 \textit{Lower panels:}}
 Fraction of total volume occupied by voids smaller than $ {\mathrm {R_{void}^{max}}}$ 
 for voids with and without a surrounding overdense shell 
 in SDSS (solid lines, filled circles) and in the mock (dashed lines, open circles) catalogues. 
 }
 \label{F_SDSS_vol} \end{figure}


\section{Large scale regions surrounding voids in galaxy catalogue} \label{S_modes}

\citet{sheth_hierarchy_2004} proposed that void evolution is strongly
determined by the larger regions surrounding them.
In this framework, voids embedded in overdense environments tend to
experience gravitational collapse more likely than expansion.
Consequently, it is expected that most of the small voids with an overdense 
shell surrounding them have sank inward by the present epoch.
Larger voids, on the other hand, are probably expanding in concordance
with the formation of large structures shaping them.
We analyze the relative abundance of voids embedded in overdense
environments for different void sizes in the galaxy distribution.
Such abundances are used in this section to test model predictions 
about voids in the SDSS spectroscopic galaxy catalogue.

\subsection{Void Density}

\citet{padilla_spatial_2005} provide mean relations
between the maximum density, its radial distance to the void centre,
and the void radius.
In order to further explore these relations we study
the integrated density contrast profiles of galaxies as a function of
void radius and distance to the void centre.
We compute for each void the integrated number density contrast of
galaxies, $\Delta_g$, as:

\begin{equation}
  \Delta_g (\mathrm d) = 
  \frac{\mathrm{n_g(d)}-\langle \mathrm{n} \rangle}{\langle \mathrm{n}
  \rangle},
\end{equation}

\noindent
where $\langle \mathrm{n} \rangle$ is the mean number density of galaxies 
in the catalogue and $\mathrm{n_g(d)}$ is the number density of galaxies
within a void--centered sphere of radius $d$.
In order to compensate for the effects due to the survey limits and geometry  %
on our density calculations we use   %
appropriate random catalogues to estimate densities.  %

In the left panel of Figure \ref{F_SDSS_dens} we show  $\Delta_g$
as a function of void-centric distance (d) for all voids
with size ($\rvoid$) in the range \mbox{6-8 h$^{-1}$Mpc} (grey solid
lines).
The thick black solid line represents the mean integrated density contrast
for this void subsample, hereafter  $\langle\Delta_g\rangle$.
By inspection of this panel, it can be
noticed that there is a significant variation of individual curves
around the mean.
We also highlight with thick dashed lines two profiles
that illustrate two distinct behaviours: 
profiles with a noticeable maximum density contrast and a decline at
larger distances; and those showing a non-decreasing profile
that tends to the mean density at large distances from the void centres.

 \begin{figure}
 \includegraphics[width=84mm]{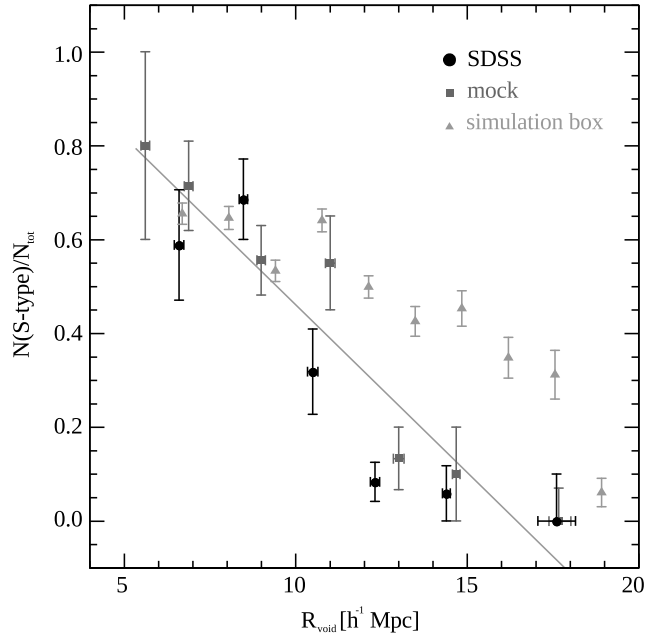}
 \caption{ 
    Fraction of S-Type voids as a function of void radius.
    Filled dots represent the fractions for the SDSS sample,
    grey boxes the corresponding fractions obtained in the mock sample
    and grey triangles the values computed in the simulation box, in
    real space.
    The solid line indicates the linear regression fit to SDSS data.
    Error bars represent the 68\% confidence interval of the binomial distribution.}
 \label{F_SDSS_frac} \end{figure}

In the right panel of Figure \ref{F_SDSS_dens}, we show
$\langle\Delta_g\rangle$ as a function of $\mathrm{(R_{void},d)}$ for
voids with radii in the range $5<\rmpc<16$.
An element in this matrix represents $\langle\Delta_g\rangle$,
averaged over voids in a given void size bin, integrated up to a given
maximum void-centric distance d.
Therefore, columns in this matrix are the mean integrated density
contrast profile as a function  of void size and projected distance.
The superimposed dashed lines represent different density contrast levels.
The solid line corresponds to a fit to the zero density contrast
iso-contour in the range $5<\rmpc<16$, shown for future reference (see
Section \ref{S_teo}).
Underdense and overdense regions are shown with different colours,
ranging from blue to orange for negative to positive density
contrasts, respectively.
The white color corresponds to a zero density contrast, i.e., regions where
the integrated density is equal to the mean global density.
This panel shows the general trend of the average behaviour of void
profiles depending on void sizes.
In the mean, the integrated density contrast profile of small voids
exhibits a prominent shell, whereas in the case of large voids there is
a smooth behaviour towards the global mean galaxy density.

The results shown in the left panel of Figure \ref{F_SDSS_dens} suggest that it is
possible to classify voids according to their large-scale radial
density profiles allowing for a subdivision of the void sample into two
types of voids.
We notice that mean integrated density contrast profiles can be
defined for $\rvoid$ intervals.
These average curves have a well defined maximum at a distance
$\mathrm {d_{max}}$ from the void centre, except for the largest voids
that exhibit an asymptotically increasing profiles.
When there is not a clear local maximum, we adopt $\mathrm {d_{max}}=3
\rvoid$. This choice is justified by the fact that small and
intermediate size voids have $\mathrm {d_{max}}  \sim 3
\rvoid $ (right panel of Figure \ref{F_SDSS_dens}).
Thus, we classify voids into two subsamples according to
positive or negative values of the integrated density contrast at
$\mathrm{d_{max}}$.
Voids surrounded by an overdense shell (hereafter S-Type voids)
correspond to $\Delta_g(\mathrm {d_{max}}) > 0$.
On the other hand R-Type voids ($\Delta_g(\mathrm{ d_{max}}) <
0$) correspond to voids with continuously 
rising density profiles.

In the left panels of Fig. \ref{F_SDSS_prof} we show the galaxy
density profile for S-Type voids with radii in two ranges $6$
h$^{-1}$Mpc $< \rvoid < 8$ h$^{-1}$Mpc and $10$ h$^{-1}$Mpc $<
\rvoid < 12$ h$^{-1}$Mpc.  The solid lines correspond to the mean
density whereas dots correspond to the individual void profiles. 
The profiles of R-Type voids are shown in the right panel of this
figure.
It can be appreciated the similarity of the inner radial density
profiles of S and R-Type voids, regardless of their different large
scale environment.
For the total sample of voids, we obtain 59 S-Type and 82 R-Type
voids.
The solid lines in the upper panels of Figure \ref{F_SDSS_vol} show the 
cumulative number of voids corresponding to the S-Type  (left panel) 
and R-Type sample (right panel) in the SDSS catalogue.
It can be noticed the opposite trends distinguishing both samples:  
while the quantity of S-Type voids decreases as the radius increases, 
that of R-Type voids monotonically increases.
This is consistent with the theoretical results indicating 
that large voids are unlikely to be surrounded by overdense shells 
\citep{more_a_do_about_nothing}.

The relative volume occupied by voids smaller than
a given maximum void radius
$\mathrm{R_{void}^{max}}$ is shown in the lower panels of Figure
\ref{F_SDSS_vol}, for S and R-types separately.
As it can easily be seen in the left lower panel, approximately 10$\%$
of the catalogue is occupied by S-type voids (solid lines, filled
dots).   
On the other hand, the volume occupied by R-Type voids increases with
radius (solid lines and filled dots in the lower right panel of Figure
\ref{F_SDSS_vol}) and they approximately occupy half of the total
volume, in agreement with the well known statement that most of the
volume of the Universe is filled by voids.

\begin{figure*}
\centering
\includegraphics[width=0.45\textwidth,height=0.45\textwidth]{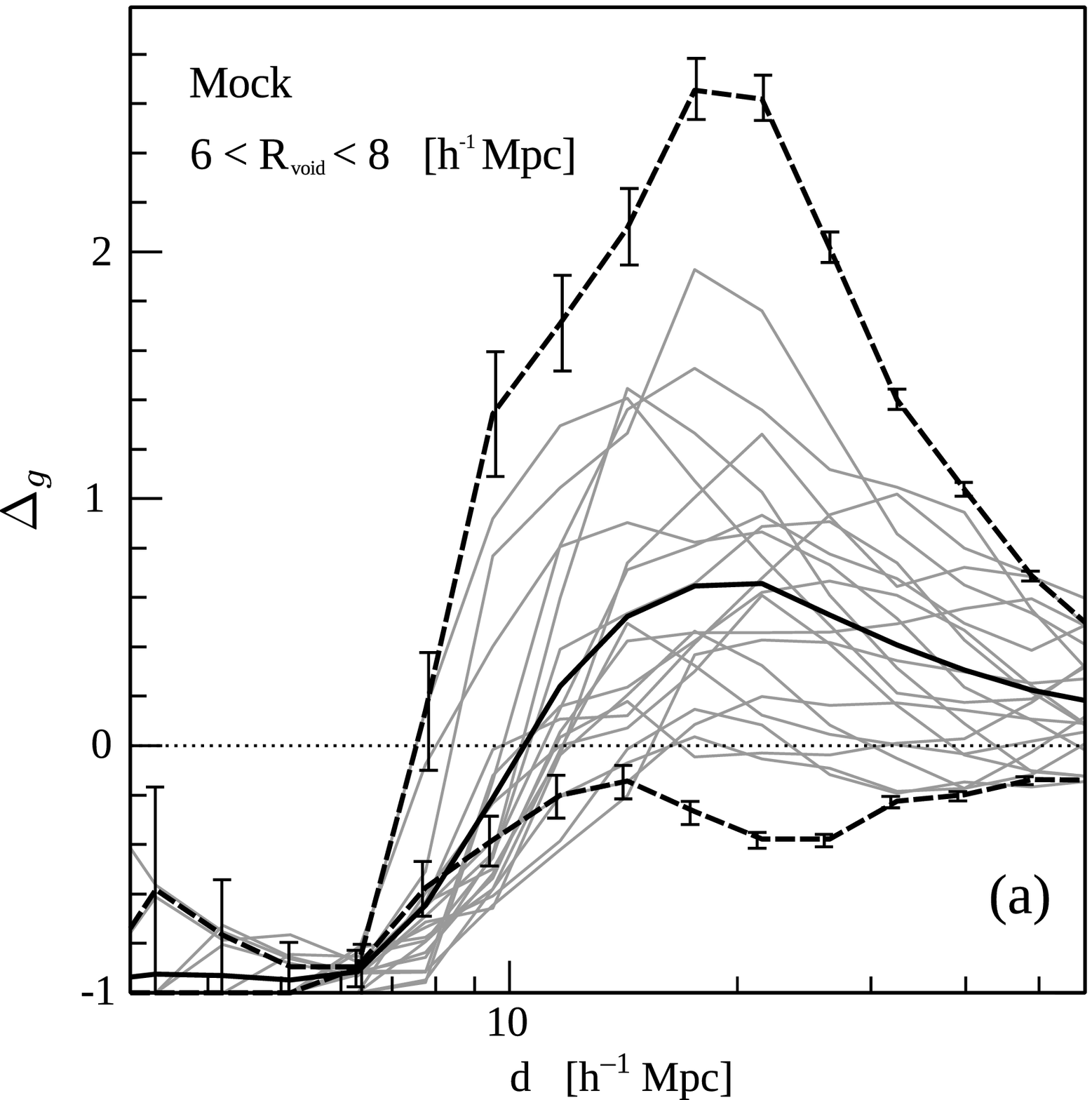}
\hspace{12pt}
\includegraphics[width=0.45\textwidth,height=0.45\textwidth]{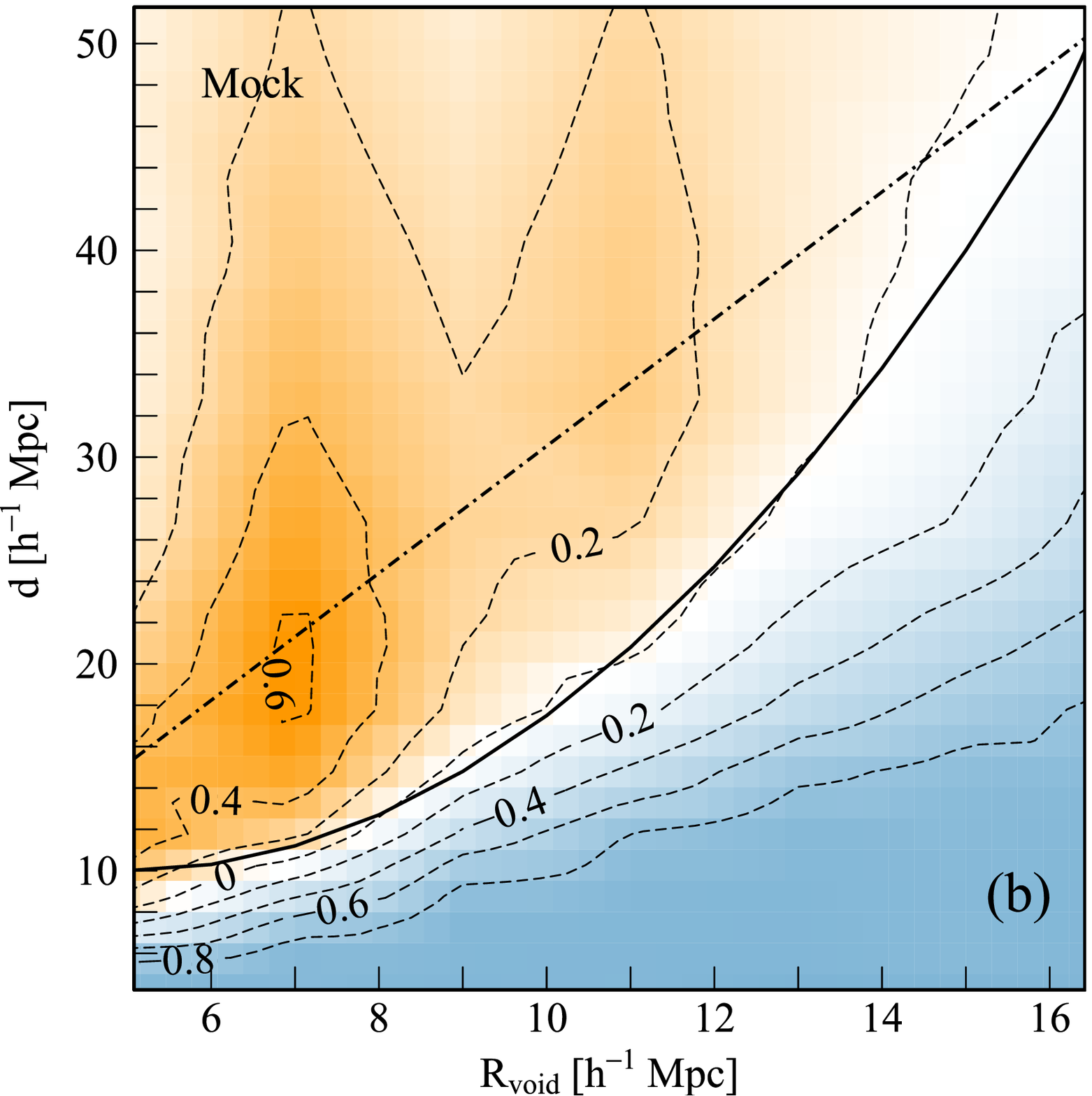}

\vspace{12pt}

\includegraphics[width=0.45\textwidth,height=0.45\textwidth]{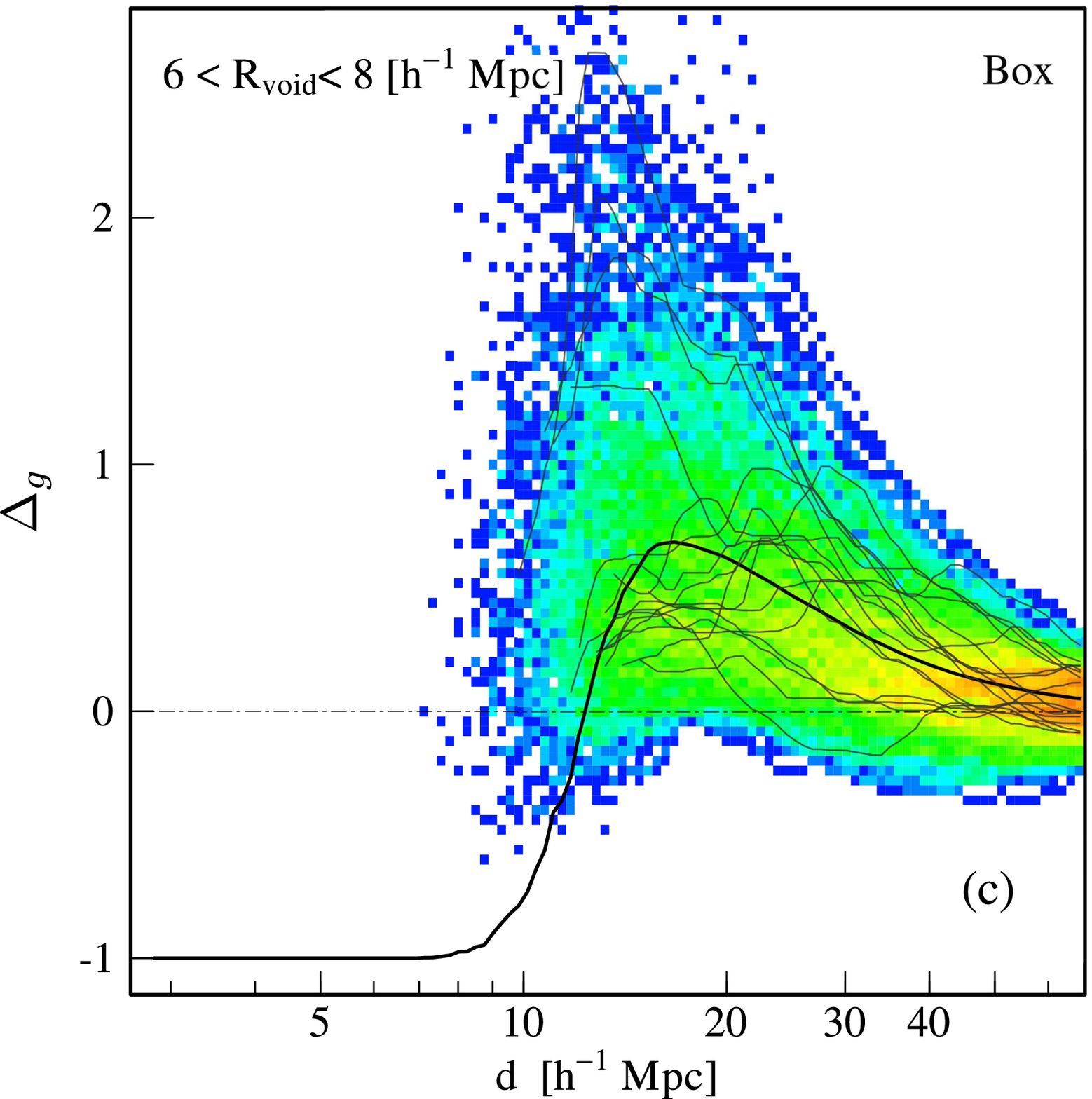}
\hspace{12pt}
\includegraphics[width=0.45\textwidth,height=0.45\textwidth]{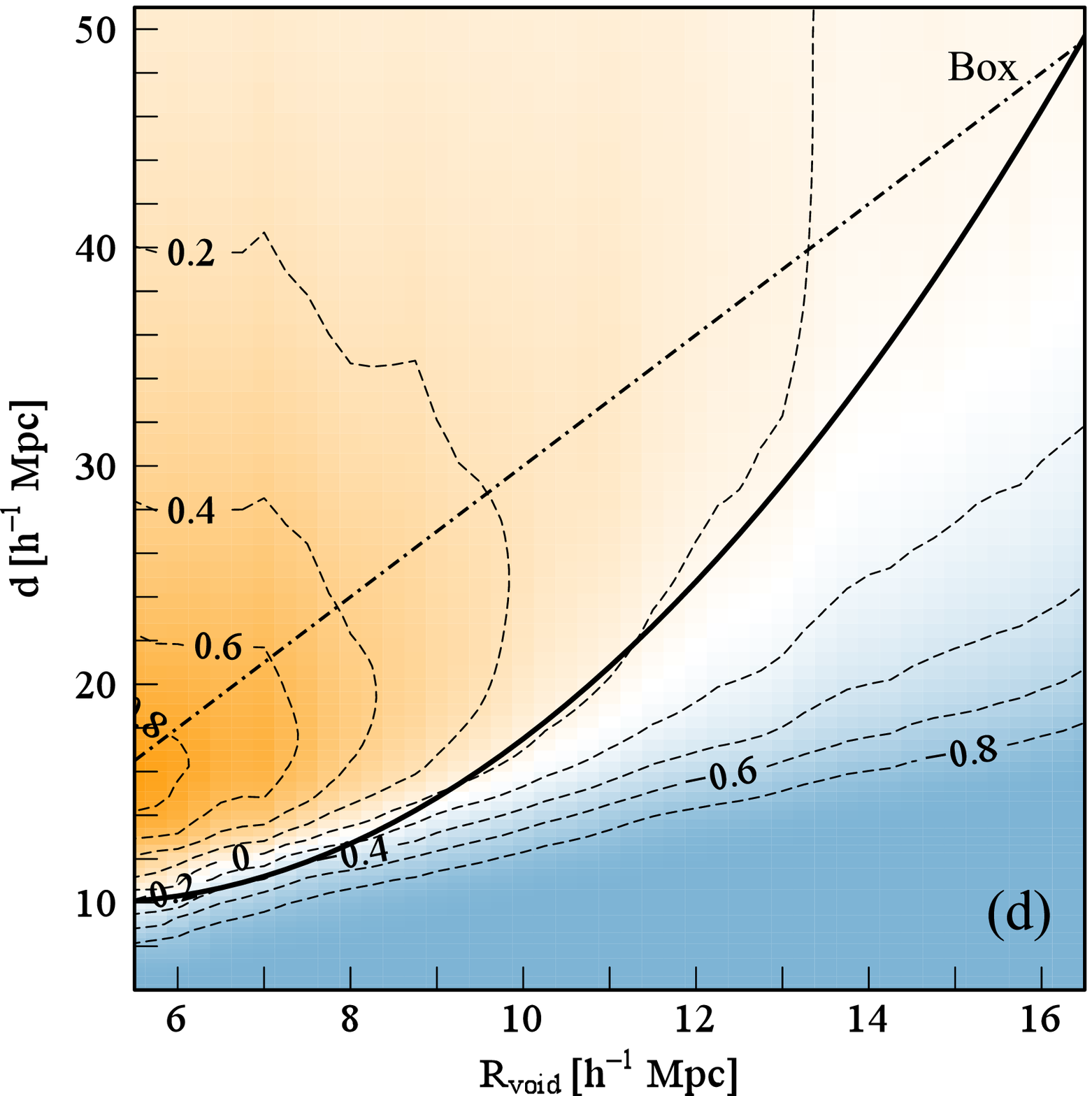}
\caption{
\textit{
	\textbf{
Left panels}} (a and c):
Integrated galaxy density profiles ($\Delta_g$, gray lines) of individual voids
with sizes between 6 and 8 h$^{-1}$Mpc,
as a function of distance to the void centre ($d$) for voids in the mock
catalogue (panel a) and in a random subset of voids in the simulation
(panel c).
The black solid lines represent the mean integrated density profile, $\langle\Delta_g\rangle$,
in both panels.
The black dashed lines in panel (a) 
correspond to the profiles of two voids
that are classified clearly into S and R types. The error bars indicate 
the corresponding poisson errors. 
The color map in panel (c) is the
histogram of the number of voids in bins of integrated galaxy
density ($\Delta_g$) and void centre distance ($d$)
(redder colors for higher counts).  
\textit{
	\textbf{
Right panels}} (b and d):
Contour lines of mean density, $\langle\Delta_g\rangle$, as a function of the void radius
$\rvoid$, and the distance to the void centre $d$, in the mock
catalogue (panel b) and in the simulation box (panel d).
Orange colours represent positive densities (increasing from white)
and cyan correspond to negative densities.  Dashed lines represent
isodensity contours (-0.8, -0.6, -0.4, -0.2, 0, 0.2, 0.4, 0.6, 0.8).
Solid lines represent an approximation
to the $\Delta_g=0$ isodensity contour in the mock sample, 
and the dot-dashed lines indicates the relation 3$\times \rvoid =$ d, 
also shown for comparison in panel (d). 
}
\label{F_MOCK_dens}
\end{figure*}

\subsection{Fraction of S and R-Type voids}

In this subsection we analyse the fraction of S and R-Type voids as a
function of void radius.
The results are displayed in Fig. \ref{F_SDSS_frac} for SDSS data and
the results of the mock catalogue (discussed in section \ref{S_teo}).
Error-bars correspond to the 68\% confidence interval of the binomial ditribution.
It can be seen in this figure that most of the smallest voids are
surrounded by an overdense region while larger voids tend to
have smoothly rising density profiles. 
A similar behaviour has been reported by
\citet{sheth_hierarchy_2004}, based on a void sample identified in
numerical simulations.
They provide a theoretical framework based on the excursion set
formalism where the void population is twofold, according to their
evolutionary processes.
Our classification of voids density profiles resembles the idea of a
void-in-void and void-in-cloud dichotomy in the evolutionary processes
presented by \citet{sheth_hierarchy_2004}.
We have found that the relative number of voids in S-Type decreases as
the size increases.
The comparison with simulation results is detailed in the next
section.

\section{$\Lambda$CDM comparison} \label{S_teo}

In this section we use the mock catalogue constructed using the numerical
simulation and identify and classify voids using the same methods as
applied to the observations in the previous sections.
The Millennium simulation, described in Section \ref{S_data}, provides
three dimensional positions and velocities of semi--analytic galaxies.
This allows to analyze the properties of voids defined by galaxy positions
for different types of large-scale
surrounding regions.
The simulation and the mock catalogue can also be used to test
possible systematics in the definition and properties of voids, which
can change from real to redshift space \citep{schmidt_size_2001,
ryden_voids_1996}.

With the aim to compare the results of galaxy catalogue to numerical
results, we study the behaviour of the void integrated density
profiles $\Delta_{g}(d)$ as a function of both void radius
($\rvoid$) and void--centric distance ($d$), in the mock catalogue
and in the simulation box.
The semi--analytic sample of galaxies comprises all galaxies from the
\mbox{$500 h^{-1}$Mpc$^3$} simulation box.
As is described in subsection \ref{ss:sim_mock}, the mock galaxy
catalogue is constructed from the semi--analytic sample of galaxies,
and with the same angular mask and redshift space effects than the
SDSS spectroscopic galaxy catalogue.
This allows a better understanding of the spread in the behaviours of
the different void profiles.
In panel (a) and (c) of Fig.  \ref{F_MOCK_dens} we show the integrated
galaxy overdensity profiles of voids with sizes ranging \mbox{6-8
h$^{-1}$ Mpc} in both, the mock catalogue and the simulation box.
In order to avoid shot--noise in the estimation of the density profile, 
we require galaxy counts to be above 100 in each radial distance bin.
Due to this limitation, the curves are trimmed at small radial
distances.
As can be seen in panel (a), the individual curves span a wide range
of behaviours for the galaxy number overdensity $\Delta_g$ as a
function of centre void distance $d$.
Similar features are observed, although with better coverage, in the
sample of voids in the simulation box (panel c of Fig.
\ref{F_MOCK_dens}). Given the high number of voids found in the
simulation volume we show just a random subset of void profiles (gray
lines).
The curves can again be subdivided in groups showing two behaviours,
that of a prominent maximum indicating the presence of an averaged
overdensity in a surrounding shell, and the characteristic smooth rise
to $\Delta=0$ at large distances.
In order to illustrate these distinct behaviours, the black dashed
lines in panel (a) show the profiles of two voids that are clearly
classified into S and R types.
The black solid line in both panels, a and c, represents the mean
integrated density profile $\langle\Delta_g\rangle$ of the mock and
simulation samples, respectively.
In the case of the simulation box, due to the larger volume covered,
the number of voids in the sample is larger, allowing a more detailed
analysis of the behaviour of void profiles. 
For instance, in panel (c) the color map represents a 2-D histogram of
the number of voids in bins of the integrated galaxy density profile
($\Delta_g$) and the void--centric distance ($d$), where redder
colours indicate higher number counts, and blue colour accounts for
one measurement in that bin.
As can be seen, the lack of red coloured
bins at separations $<30$h$^{-1}$Mpc shows that the distribution of
overdensity values becomes broader.
It should be noticed that each curve is sampled at a discrete set of
values for the void-centric distance (the bins in $d$), giving rise to
the noisy aspect of the colour map for low density values.

In the above paragraph we have illustrated for a fixed range of void sizes
(\mbox{6-8 h$^{-1}$ Mpc}) the features observed in void profiles.
We also provide in figure \ref{F_MOCK_dens} an analysis of the mean void
density profile $\langle\Delta_g\rangle$ as a function of void size.  
This is shown in panels (b) and (b), where mean galaxy overdensity is presented
as a function of the void radius ($\rvoid$) and the distance to the void
centre ($d$) in the mock catalogue and in the simulation box, respectively.
Orange colours represent overdense regions ($\langle\Delta_g\rangle > 0$)
whereas cyan colours correspond to underdense regions (negative overdensities). 
The dashed lines represent isodensity contours as labeled in the figure.
The Solid lines in both panels represent an approximation to the $\Delta_g=0$
isodensity contour in the mock sample, also shown for comparison in the
simulation sample (panel d).
As it can be noticed, small voids tend to exhibit a surrounding overdense shell
in their mean profile, whereas larger voids \mbox{($10 \lesssim \rvoid$)} show
continuously rising mean density profiles.
It is worth to mention that all the observational effects taken into account in the construction of the
mock catalogue do not seem to affect in an appreciable way the mock results.
This arises from comparing the mock results (upper panels in figure
\ref{F_MOCK_dens}) with the simulation results (lower panels).

In order to examine cosmic variance effects on our results,
we have also constructed several mock catalogues by placing observers in unconnected
locations in the simulation box. We examine the density profiles around voids. 
In general they show similar behaviour with small differences between 
different mocks which are comparable to the difference between mock and 
observational data (see Figures \ref{F_SDSS_dens} and \ref{F_MOCK_dens}).%
We stress the fact that 
our theoretical results are in good agreement with%
observations as can be seen when comparing to Figure \ref{F_SDSS_dens}.

\begin{figure}
 \includegraphics[width=84mm]{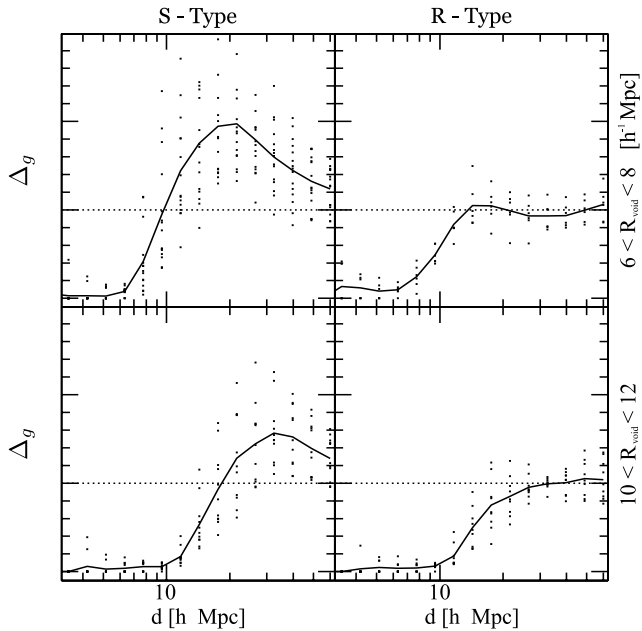}
 \caption{
    Radial integrated density profile around voids in the Mock
    catalogue
    with void radii in the ranges 6-8 h$^{-1}$Mpc (upper panels) and
    10-12 h$^{-1}$Mpc  (lower panels). 
    The left panels correspond to voids surrounded by large scale
    overdense shells and the right panels correspond to voids in large scale
    underdense regions.  
    Solid lines correspond to the mean density and dots show the
    individual voids.  Dotted lines indicate the
    mean galaxy density.  
  }
 \label{fig:prof_mock} \end{figure}

We select S and R-Type void samples in the mock catalogue following
the same criteria described in section \ref{S_modes} obtaining 48 and
65 S and R-Type voids respectively.
The density profiles corresponding to S and R-Type voids are shown in
Figure \ref{fig:prof_mock} (left and right panels respectively).
In Figure \ref{F_SDSS_frac} we present the fraction of S and R-Type
voids with respect to the total number as a function of void size in
the mock catalogue, where the squares indicate the fraction and error-bars 
represent the confidence interval of the binomial distribution.
It is relevant that the mock behaviour is indistinguishable from the
observations. 
We also show the results obtained selecting galaxies brighter than 
$M_r =$ -19.2 
in the simulation box 
(triangles in figure \ref{F_SDSS_frac}).
Thus, this sample of simulated galaxies have the same absolute limiting magnitude 
than the volume-complete samples of SDSS and mock galaxies.

\begin{figure}
 \includegraphics[width=84mm]{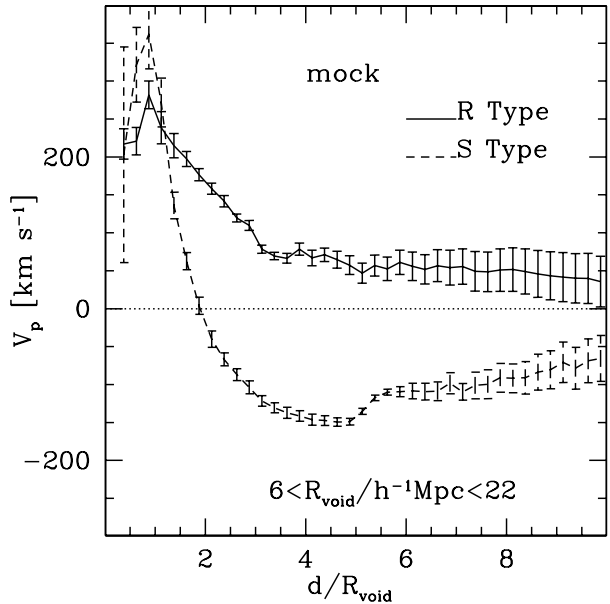}
 \caption{
  Mean radial velocity profile
  for voids surrounded (dashed line) and not surrounded (solid line) by overdense
  shells in the mock catalogue. Void radii are in the range 6-22 h$^{-1}$ Mpc.
  The dotted line indicates a  zero mean velocity (v = 0 km s$^{-1}$).
  }
 \label{F_MOCK_vel} \end{figure}

Based on analytical formulations for the evolution of inhomogeneities
on the mass distribution in the Universe
\cite[e.g.][]{peebles_principles_1993} and the theoretical analysis of
void evolution \citep{sheth_hierarchy_2004} it is natural to expect a
dependence of the peculiar velocity field around voids with the
presence of a surrounding over-dense shell. 
In order to examine this effect, we have studied the mean peculiar
velocity around voids traced by the semi-analytic galaxies in the
full simulation box, for S and R-Type voids separately.
In Figure \ref{F_MOCK_vel} we show the mean velocity profiles of
these two subsamples of voids in the mock catalogue, where the dashed 
line indicates S-Type and the solid line R-Type voids. 
We adopt positive velocities to indicate expansion and 
negative velocities for infall motions. 
As can be seen in the figure, S-Type voids show substantial infall
velocities (raising to \mbox{-150 km s$^{-1}$}) at the same distances where the 
overdense shells are located (d/$\rvoid \sim$ 3) whereas at smaller distances 
(d/$\rvoid \lesssim$ 2) the velocity field is characterized by expansion. 
On the other hand, R-Type voids, lacking an overdense shell, show only
expansion velocities.
Both samples show significant expansion velocities 
at void shells (0.8 $\leq$ d/$\rvoid \leq$ 1.2), with a 
maximum (velocities $\sim$ 300 km s$^{-1}$) reached at d/$\rvoid \simeq$1.  

%
%


\section{Summary and discussion}  \label{S_concl}

We have performed a statistical study of the void phenomenon focussing 
on void environments.
We have examined the distribution of galaxies around voids in the
SDSS, by computing the integrated density contrast profile.
There is a correlation between void--centric distance to the shell of
maximum density and void radius.
We defined separation criterion to characterize voids according to their
surrounding environment, giving rise to S-Type (Shell) and R-Type (Rising) voids.
We found that small voids are more frequently surrounded by
over-dense shells.
On the other hand, larger voids are more likely classified as
R-Type, i.e., with a non--decreasing integrated density contrast
profile, which smoothly rises towards the mean galaxy density.
The fraction of voids surrounded by overdense shells
continuously decreases as the void size increases.

In order to test and interpret the observed properties of voids we
identified and analyzed voids in a numerical simulation with a
semi-analytic galaxy catalog.  We used the same procedures than those
applied to the observational galaxy catalog to identify and analyze voids,
particularly when using the mock catalog.    
We have computed the velocity curves for the two types of voids, in
the full box and in the mock sample.
The results suggest that there is a relation between our separation
criterion and the evolution of voids, as was suggested previously by
\citet{sheth_hierarchy_2004}.
The good agreement obtained between the SDSS voids sample and the
mock catalog provides additional support to the viability of a $\Lambda$CDM
model to reproduce the large scale structure of the universe as
defined by the void network.
In this work we give the first observational evidence of multiple
modes in the void hierarchy, which have been predicted in numerical
simulations, see for example \citep{more_a_do_about_nothing} and 
references therein.


\section*{Acknowledgments}
We would like to thank the referee, Paul Sutter, for the thorough, 
constructive and helpful comments and suggestions on the manuscript, 
which greatly improved this work.
This work has been partially supported by Consejo de Investigaciones
Cient\'{\i}ficas y T\'ecnicas de la Rep\'ublica Argentina (CONICET)
and the Secretar\'{\i}a de Ciencia y T\'ecnica de la Universidad
Nacional de C\'ordoba (SeCyT). 
LC, DP and ML
acknowledge research fellowships from CONICET.
NP acknowledges support from Fondecyt 1110328 and BASAL-PFB06 "Centro de Astronom\'\i a y Tecnolog\'\i as afines".

Funding for the SDSS and SDSS-II has been provided by the Alfred P.
Sloan Foundation, the Participating Institutions, the National Science
Foundation, the U.S. Department of Energy, the National Aeronautics
and Space Administration, the Japanese Monbukagakusho, the Max Planck
Society, and the Higher Education Funding Council for England. The
SDSS Web Site is http://www.sdss.org/.
The SDSS is managed by the Astrophysical Research Consortium for the
Participating Institutions. The Participating Institutions are the
American Museum of Natural History, Astrophysical Institute Potsdam,
University of Basel, University of Cambridge, Case Western Reserve
University, University of Chicago, Drexel University, Fermilab, the
Institute for Advanced Study, the Japan Participation Group, Johns
Hopkins University, the Joint Institute for Nuclear Astrophysics, the
Kavli Institute for Particle Astrophysics and Cosmology, the Korean
Scientist Group, the Chinese Academy of Sciences (LAMOST), Los Alamos
National Laboratory, the Max-Planck-Institute for Astronomy (MPIA),
the Max-Planck-Institute for Astrophysics (MPA), New Mexico State
University, Ohio State University, University of Pittsburgh,
University of Portsmouth, Princeton University, the United States
Naval Observatory, and the University of Washington.

The Millennium Simulation databases used in this paper and the web
application providing online access to them were constructed as part
of the activities of the German Astrophysical Virtual Observatory.

Some of the plots presented in this work were made by using R Software.

\bibliographystyle{mn2e}
\bibliography{draft1.bib}

\end{document}